\documentclass[journal,twoside,web]{IEEEtran}
\usepackage{cite}
\usepackage{amsmath,amssymb,amsfonts}
\usepackage{algorithmic}
\usepackage{graphicx}
\usepackage{textcomp}
\usepackage{booktabs}
\usepackage{multirow}

\usepackage{bm}

\def\BibTeX{{\rm B\kern-.05em{\sc i\kern-.025em b}\kern-.08em
    T\kern-.1667em\lower.7ex\hbox{E}\kern-.125emX}}
\begin{document}
\title{Learning Global and Local Features of \\Normal Brain Anatomy for \\Unsupervised Abnormality Detection}
\author{Kazuma Kobayashi, Ryuichiro Hataya, Yusuke Kurose, Amina Bolatkan, Mototaka Miyake, \\Hirokazu Watanabe, Masamichi Takahashi, Jun Itami, Tatsuya Harada, and Ryuji Hamamoto

\thanks{This study was supported by JST CREST (Grant Number JPMJCR1689), JST AIP-PRISM (Grant Number JPMJCR18Y4), and JSPS Grant-in-Aid for Scientific Research on Innovative Areas (Grant Number JP18H04908). \emph{(Corresponding author: Kazuma Kobayashi.)}}
\thanks{Kazuma Kobayashi, Amina Bolatkan, and Ryuji Hamamoto are with the Division of Molecular Modification and Cancer Biology, National Cancer Center Research Institute, 5-1-1 Tsukiji, Chuo-ku, Tokyo 104-0045, Japan (e-mail: kazumkob@ncc.go.jp; abolatka@ncc.go.jp; rhamamot@ncc.go.jp). They are also with the Cancer Translational Research Team, RIKEN Center for Advanced Intelligence Project, 1-4-1 Nihonbashi, Chuo-ku, Tokyo.}
\thanks{Ryuichiro Hataya is with the Graduate School of Information Science and Technology, The University of Tokyo, 7-3-1 Hongo, Bunkyo-ku, Tokyo 113-8656, Japan (e-mail: hataya@nlab.ci.i.u-tokyo.ac.jp).}
\thanks{Yusuke Kurose and Tatsuya Harada are with the Research Center for Advanced Science and Technology, The University of Tokyo, 4-6-1 Komaba, Meguro-ku, Tokyo 153-8904, Japan (e-mail: kurose@mi.t.u-tokyo.ac.jp; harada@mi.t.u-tokyo.ac.jp). They are also with the Machine Intelligence for Medical Engineering Team, RIKEN Center for Advanced Intelligent Project, 1-4-1 Nihonbashi, Chuo-ku, Tokyo 103–0027, Japan.}
\thanks{Mototaka Miyake and Hirokazu Watanabe are with the Department of Diagnostic Radiology, National Cancer Center Hospital, 5-1-1 Tsukiji, Chuo-ku, Tokyo 104-0045, Japan (e-mail: mmiyake@ncc.go.jp; hirwatan@ncc.go.jp).}
\thanks{Masamichi Takahashi is with the Department of Neurosurgery and Neuro-Oncology, National Cancer Center Hospital, 5-1-1 Tsukiji, Chuo-ku, Tokyo 104-0045, Japan (e-mail: masataka@ncc.go.jp).}
\thanks{Jun Itami is with the Department of Radiation Oncology, National Cancer Center Hospital, 5-1-1 Tsukiji, Chuo-ku, Tokyo 104-0045, Japan (e-mail: jitami@ncc.go.jp).}
}

\maketitle

\begin{abstract}
In real-world clinical practice, overlooking unanticipated findings can result in serious consequences. However, {\emph{supervised learning}}, which is the foundation for the current success of deep learning, only encourages models to identify abnormalities that are defined in datasets in advance. Therefore, abnormality detection must be implemented in medical images that are not limited to a specific disease category. In this study, we demonstrate an {\emph{unsupervised learning}} framework for pixel-wise abnormality detection in brain magnetic resonance imaging captured from a patient population with metastatic brain tumor. Our concept is as follows: If an image reconstruction network can faithfully reproduce the global features of ``normal'' anatomy, then the ``abnormal'' lesions in unseen images can be identified based on the local difference from those reconstructed as ``normal'' by a discriminative network. Both networks are trained on a dataset comprising only normal images without labels. In addition, we devise a metric to evaluate the anatomical fidelity of the reconstructed images and confirm that the overall detection performance is improved when the image reconstruction network achieves a higher score. For evaluation, clinically significant abnormalities are comprehensively segmented. The results show that the area under the receiver operating characteristics curve values for metastatic brain tumors, extracranial metastatic tumors, postoperative cavities, and structural changes are 0.78, 0.61, 0.91, and 0.60, respectively.
\end{abstract}


\section{Introduction}
\label{sec:introduction}
\IEEEPARstart{I}{n} real-world clinical practice, overlooking unanticipated findings can have serious consequences. The recent advances in deep learning have significantly enhanced the practice and research of radiology, and one of the dominant learning frameworks is \emph{supervised learning}. Supervised learning typically requires a considerable volume of data to which paired labels have been assigned \cite{Alexnet_NIPS2012_4824}; in other words, these labels must be specifically defined in advance. Furthermore, because the costs associated with expert annotation are high, a dataset that encompasses various diseases is difficult to build. In fact, disease phenotypes are limitless because individually different genetic, environmental, and therapeutic factors affect their complex etiology and clinical courses. Moreover, comorbidities and incidental diseases can cause unexpected changes. Hence, not only to reduce the cost of annotation but also to encompass various diseases with unpredictable abnormalities, abnormality detection must be realized in medical images that are not limited to certain disease categories.

To detect undefined changes in samples, an alternative learning framework, i.e., \emph{unsupervised learning}, is required. Generally, unsupervised learning techniques aim to help models identify underlying patterns in data without annotation. In medical imaging, the pattern of normal anatomy is typically reflected in a population with a small variation range. In addition, a significant class imbalance exists between normal and abnormal features, i.e., the majority of data pertain to healthy samples, and a limited number of disease samples have a wide range of phenotypic differences. Therefore, by learning normal anatomical features that recursively appear in human anatomy, abnormalities in unseen images can be identified as features deviating from them. This type of outlier detection task, which is based on certain assumptions regarding ``normality,'' should be formulated as an \emph{unsupervised abnormality (anomaly) detection}. Such tasks may appear simple; however, they are still technically challenging as machine learning problems \cite{Chalapathy2019}. It is difficult to model the distribution of normal features and distinguish abnormal features from the distribution. 

When modeling the distribution of normal anatomy, the hierarchical characteristic of abnormalities caused by diseases should be considered. Here, we divide the abnormality in medical images into global and local changes from the normal baseline. Global abnormality occurs when a structure appears in unexpected locations, including structural deviations due to surgery, trauma, degenerative changes, or compression from space-occupying lesions. Meanwhile, local abnormality appears as a localized lesion that has been replaced by a disease, such as tumor, vascular disease, and inflammation. Occasionally, a single disease is accompanied by both global structural changes and local abnormalities. For example, brain tumors appear as focal changes, and with their enlargement, cause peritumoral edema and compressional deformation of adjacent structures. Therefore, physicians combine global and local features when diagnosing whether an image of interest is abnormal. 

In this study, we devise an algorithm for detecting a wide variety of abnormalities without labels, which appear globally or locally in medical images, and report its diagnostic performance based on a real-world dataset.

\subsection{Proposed Methods}
The proposed algorithm utilizes a two-step approach to account for both global and local features of medical images. It comprises an image reconstruction network that reproduces the global features of normal anatomy from input images and a discriminative network to identify local deviations of imaging features from reconstructed references. Both networks are trained using only normal images without labels, i.e., the former self-learns a latent distribution of image-level normal anatomy, whereas the latter self-learns a latent distribution of patch-level normal anatomy. 

A faithful reconstruction of normal anatomy on an image-by-image basis is required in the image reconstruction network. Hence, we first introduce a variational autoencoder (VAE) \cite{Kingma2013, Rezende2014}. A VAE is effective for modeling the distributions of high-dimensional data through variational inference. Our basic concept is to train a VAE using a series of images without abnormalities such that a latent distribution $p(\bm{z})$ representing the normal anatomy can be acquired inside the model. Subsequently, the trained model can map an unseen image to a point on the manifold that represents the closest normal representation to them and then reconstruct a normal-appearing replica in the image space. Furthermore, to achieve a successful image reconstruction with sufficient fidelity, we employed an \emph{introspective VAE (IntroVAE)} \cite{Huang2018}, which is an extended framework of the VAE. By introspectively self-evaluating the differences between the input and reconstructed images, IntroVAE can synthesize more realistic, high-resolution images. In addition, we applied a \emph{latent representation search} to obtain a better latent representation for the input images in order to achieve more accurate reconstructions. 


Using normal-appearing images in the reconstruction network as a reference, we aim to establish a method to localize abnormal findings in unseen images. Although per-pixel differences between the query and reconstructed images are widely used \cite{BAUR2021101952}, it might be difficult to suppress noisy information due to unimportant differences in image details. Therefore, the discriminative network, which was trained using metric learning based on abnormality-free samples, was employed to calculate the \emph{abnormality score} based on the patch-wise similarity of embedded features. The proposed method, utilizing the abnormality score, yielded a well-corresponding accumulation in the presence of unseen semantic objects (i.e., abnormal findings).


In addition, we propose \emph{anatomical fidelity} for the quantitative evaluation of faithfulness. The basic concept is that the generated image should exhibit not only high-resolution details at local points, but also anatomical consistency between distant portions of the images. We decomposed this index into two measures by using a multiclass segmentation network trained on the same image domain, which we refer to as the \emph{quality score} and \emph{overlap score}. The quality score tends to be high if the generated images contain clear objects that are sharp rather than blurry. In this case, the segmentation network can perform pixel-wise classification with high confidence, accompanied by low entropy. The overlap score is calculated based on the overlap between the segmentation results for several anatomical classes.

To evaluate the performance of diverse abnormalities appearing in real clinical practice, such as the various imaging phenotypes of a disease and post-treatment changes, we prepared a large brain magnetic resonance imaging (MRI) dataset obtained from a population with metastatic brain tumor. Owing to the diversity of primary tumors with different biological backgrounds, the imaging appearance of brain metastasis shows a wide range of phenotypes in a population. Moreover, brain MRIs with at least one brain metastasis can contain various structural abnormalities, some of which are related to disease progression, while the others are treatment-induced. In this study, to create a dataset that comprehensively spans the possible structural abnormalities, each MRI volume was segmented into four classes: \emph{metastatic brain tumor}, \emph{extracranial metastatic tumor}, \emph{postoperative cavity}, and \emph{structural change, not otherwise specified (NOS)}. These fine-grained definitions of abnormality provide an opportunity to interpret the model performance for real-world clinical practice. 

In summary, the main contributions of the present study are as follows:
\begin{itemize}
 \item We demonstrate a simple, but effective, two-step approach to detect abnormalities appearing globally or locally in medical images by leveraging an image reconstruction network followed by a discriminative network, both of which are trained without labels to learn normal anatomical representation; 
 \item We quantitatively evaluate the anatomical fidelity (an index incorporating both the image quality and anatomical consistency) of the images generated via image reconstruction networks and demonstrate a positive relationship with the overall detection performance of abnormality; 
 \item We evaluate the detection performance of the model based on the dataset, comprehensively including the structural abnormality of the brain parenchyma; hence, the utility for identifying unexpected abnormalities that appear in actual clinical practice can be discussed.
\end{itemize}

\subsection{Related Studies}

Many approaches for unsupervised abnormality detection in medical image analysis have been proposed over the last two decades. These approaches employed stochastic intensity models {\cite{VanLeemput2001}}, atlas-based registration {\cite{Prastawa2004}}, clustering {\cite{Shiee2010}}, content-based retrieval {\cite{Weiss2013}}, and statistical models based on image registration {\cite{Tomas-Fernandez2015, Zeng2016}}. More recently, advancements in computer vision have enabled the development of deep-learning-based models for unsupervised abnormality detection {\cite{An2015VariationalAB, Chalapathy2017}}. Herein, we describe relevant deep-learning-based models for pixel/voxel-wise abnormality detection in medical images, focusing on reconstruction-based and discriminative-boundary-based approaches. It is noteworthy that our method combines both approaches; the former pertains to the image reconstruction network, whereas the latter conceptually corresponds to the discriminative network. 

\subsubsection{Reconstruction-based Approaches} 

Deep generative models, such as generative adversarial networks (GAN) \cite{Goodfellow2014} and VAEs \cite{Kingma2013, Rezende2014}, assume a low-dimensional latent distribution $p(\bm{z})$ for a specified data distribution $p_\mathrm{data}$. When trained only on healthy samples, the latent distribution can represent the variation in normal anatomy. Hence, the residual errors between an input and a reconstructed image, which is generated by mapping a latent point back to the original space, can be informative for abnormality detection. Schlegl et al. \cite{Schlegl2017} first proposed AnoGAN, which utilizes a deep convolutional GAN \cite{radford2015unsupervised} trained on healthy optical coherence tomography (OCT) images of the retina. Subsequently, Schlegl et al. \cite{Schlegl2019} presented fast AnoGAN, which is a modified version of AnoGAN, enabling the fast mapping of query images to the latent space. Meanwhile, other generative models, such as Bayesian autoencoders \cite{Pawlowski2018MIDL}, VAEs \cite{Chen2018MIDL}, adversarial autoencoders \cite{Chen2018MIDL}, and Gaussian mixture VAE \cite{Chen2020} have been applied for the segmentation of glioma. Recently, Baur et al. \cite{BAUR2021101952} performed a comparative study by utilizing a single architecture, a single resolution and the same dataset for unsupervised anomaly segmentation in brain MRI, showing that different approaches exhibit different discrepancies between reconstruction-error statistics, which can be the best indicator for good detection performance.



\subsubsection{Discriminative-boundary-based Approaches}

Another approach has been proposed to seek discriminative boundaries around normal training samples in a feature space \cite{Chandola2009}. Deep neural networks can be utilized as feature extractors, which are necessary to reduce the dimensionality of high-dimensional data including images, to construct an underlying distribution of normal data. Thereafter, a discriminative model was applied to create the decision boundaries. For example, Seeb\"ock \cite{Seebock2019} et al. combined a multiscale deep denoising autoencoder and a one-class support vector machine (oc-SVM) for the unsupervised detection of age-related macular degeneration in OCT images. Recently, Alaverdyan et al. \cite{Alaverdyan2020} proposed a regularized twin neural network with oc-SVM to perform the challenging task of detecting subtle epileptic lesions in multiparametric brain MRI. 

\section{Methodology}
\label{sec:methodology}

The proposed method is composed of two stages. The first stage involves training a reconstruction network based on VAE to represent image-level normal anatomical variability, and includes additional methods to enhance the anatomical fidelity of reconstructed images. The second stage involves calculating the pixel-wise abnormality score for abnormality detection, which is obtained from the discriminative network to recognize patch-level discrepancy between input images and reconstructed images. 

\subsection{Notation}

We consider a single-modality three-dimensional (3D) MRI volume $\bm{X} \in \mathbb{R}^{C \times I \times J \times K}$, where $C$ is the number of channels; $I$ and $J$ represent the height and width of the axial slices, respectively; and $K$ is the number of axial slices. We define $\bm{x} \in \mathbb{R}^{C \times I \times J}$ as a slice in the axial view. The image reconstruction network maps the slice-wise input $\bm{x}$ into the low-dimensional latent representation $\bm{z} \in \mathbb{R}^{C^{\prime} \times I^{\prime} \times J^{\prime}}$ and reconstructs it to the image space denoted by $\hat{\bm{x}} \in \mathbb{R}^{C \times I \times J}$. The latter can be concatenated in an orderly manner to yield the corresponding volume $\hat{\bm{X}} \in \mathbb{R}^{C \times I \times J \times K}$. 


\begin{figure}
  \centering
  \includegraphics[width=\hsize]{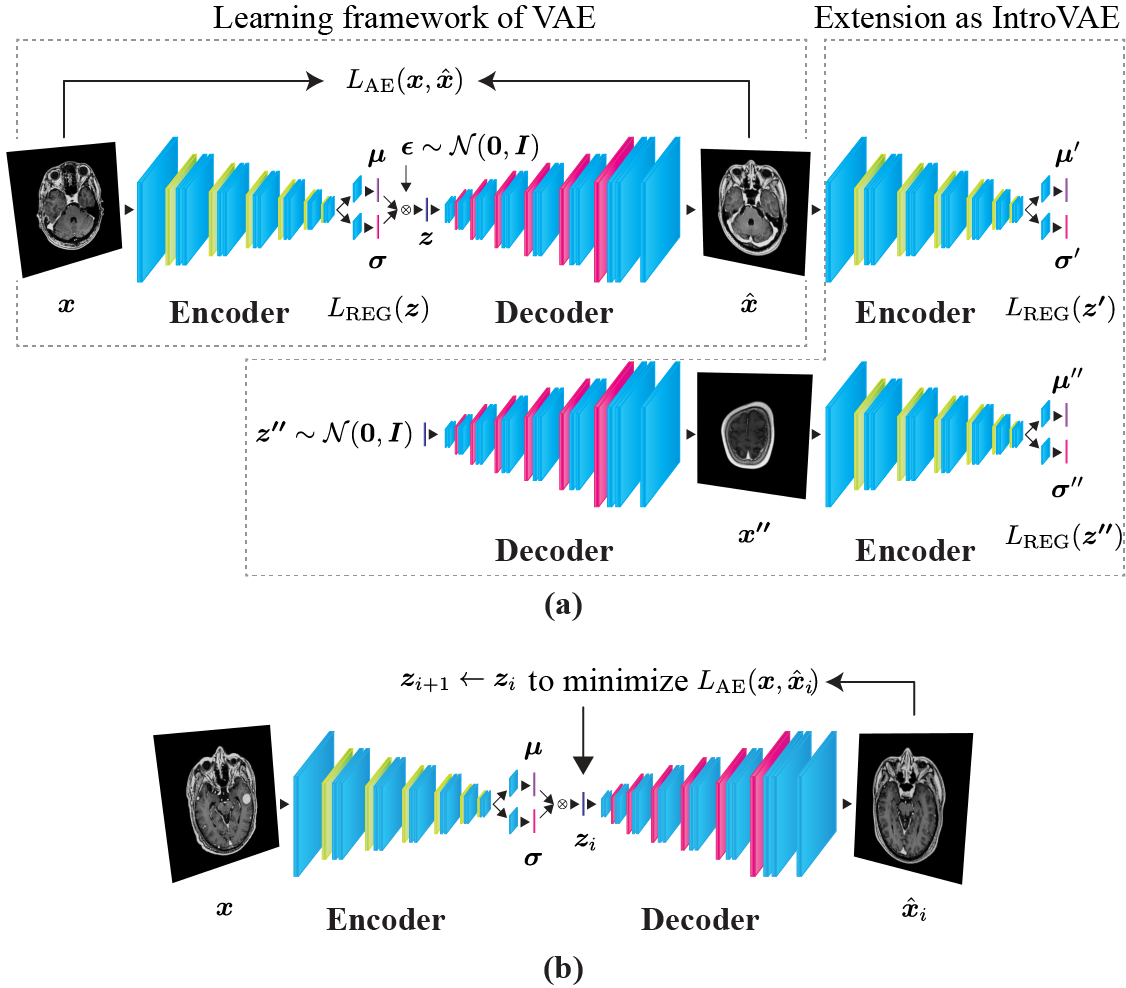}
  \caption{\textbf{Schematic illustration of image reconstruction network for learning global features of normal brain anatomy.} (a) Architecture comprising an encoder network and a decoder network. The input image $\bm{x}$ is mapped to a low-dimensional latent space through the encoder. The decoder generates its reconstruction $\bm{\hat{x}}$ from the sampled latent variable $\bm{z}$. In the learning framework of IntroVAE, the encoder determines whether the input is from the data distribution or the decoder by changing the destination of its mapping function, whereas the decoder generates more realistic images to deceive the encoder. (b) Latent representation search obtains a better latent representation $\bm{z_*}$ for a better reconstructed image $\bm{\hat{x}_*}$ with lower reconstruction error.}
  \label{fig:image_reconstruction_networks}
\end{figure}

\subsection{Image Reconstruction Network}

To construct a low-dimensional manifold representing image-level normal brain anatomy, we exploited a VAE as the basic architecture for the image reconstruction network. Subsequently, we utilized IntroVAE as an extension of the VAE as well as for latent representation searching to enhance the anatomical fidelity of the reconstructed images (Fig. \ref{fig:image_reconstruction_networks}).


\subsubsection{Introduction of VAE}

VAEs use variational inference to approximate a specified data distribution $p_\mathrm{data}$ by a latent distribution $p(\bm{z})$. They comprise a pair of encoder and decoder networks. The encoder functions as an inference network for posterior distribution $q_\phi (\bm{z}|\bm{x})$, enabling the projection of an input variable $\bm{x}$ into a corresponding latent variable $\bm{z}$. Subsequently, the decoder models the likelihood $p_\theta (\hat{\bm{x}}|\bm{z})$ by producing a visible variable $\hat{\bm{x}}$ based on the latent variable $\bm{z}$. An isotropic multivariate Gaussian $\mathcal{N}(\bm{z}; \bm{0}, \bm{I})$ is often selected as the distribution $p(\bm{z})$ over the latent variables. Hence, based on two encoder output variables, $\bm{\mu}$ and $\bm{\sigma}$, the posterior distribution is estimated to be $q_\phi (\bm{z}|\bm{x}) = \mathcal{N}(\bm{z}; \bm{\mu}, \bm{\sigma}^{2})$. Notably, the input variable for the decoder $\bm{z}$ can be sampled from $\mathcal{N}(\bm{z}; \bm{\mu}, \bm{\sigma}^{2})$ using a reparameterization trick: $\bm{z} = \bm{\mu} + \bm{\sigma} \odot \bm{\epsilon}$, where $\bm{\epsilon} \sim \mathcal{N}(\bm{0}, \bm{I})$, and $\odot$ denotes the Hadamard product. The learning objective of VAEs is to maximize the evidence lower bound of $p_{\theta}(\bm{x})$, and it can be defined as follows:
\begin{equation} 
 \log{p_\theta (\bm{x})} \geq \mathbb{E}_{q_\phi (\bm{z}|\bm{x})} \log{p_\theta (\hat{\bm{x}}|\bm{z})} - \mathrm{KL} (q_\phi (\bm{z}|\bm{x}) || p(\bm{z})),
\end{equation} 
where $\mathrm{KL}(\cdot || \cdot)$ is the Kullback--Leibler (KL) divergence between two probability distributions. The first term is a negative log likelihood, which can be proportional to the squared Euclidean distance between the input $\bm{x}$ and reconstructed $\hat{\bm{x}}$ images \cite{doersch2016tutorial}. The second term causes the approximated posterior $q_\phi (\bm{z}|\bm{x})$ to be close to the prior $p(\bm{z})$. These terms are assigned separate labels below for further consideration. 
\begin{equation} 
{L}_\mathrm{AE} = - \mathbb{E}_{q_\phi (\bm{z}|\bm{x})} \log{p_\theta (\hat{\bm{x}}|\bm{z})},
\end{equation} 
\begin{equation} 
\label{eq:regularization_term}
{L}_\mathrm{REG} = \mathrm{KL} (q_\phi (\bm{z}|\bm{x}) || p(\bm{z})).
\end{equation} 



One of the limitations of VAE is that the generated samples tend to be blurry \cite{Larsen2015, zhao2017infovae}. Because this shortcoming may hinder faithful reconstruction of input images, we explored the extension of VAE to achieve better anatomical fidelity. 



\subsubsection{Introduction of IntroVAE}

We utilized IntroVAE \cite{Huang2018}, which is an extended architecture of the VAE. By introspectively self-evaluating the differences between the input and reconstructed images, IntroVAE can self-update to synthesize more realistic, high-resolution images. It is noteworthy that a min--max game exists between the encoder and decoder, similar to that employed in GANs. The encoder is trained to determine whether the input images are from a data distribution $p_{\mathrm{data}}$ or the decoder $p_\theta(\hat{\bm{x}}|\bm{z})$, whereas the decoder is encouraged to ``fool'' the encoder by generating realistic images.

In the learning framework of IntroVAE, ${L}_\mathrm{REG}$ is extended as an adversarial cost function. The encoder is trained to minimize ${L}_\mathrm{REG}$ for real images $\bm{x}$ to match the posterior $q_\phi (\bm{z}|\bm{x})$ to the prior $p(\bm{z})$, and conversely, to increase ${L}_\mathrm{REG}$ for the generated images $\hat{\bm{x}}$ such that the posterior $q_\phi(\bm{z}^{\prime}|\hat{\bm{x}})$ deviates from the prior $p(\bm{z})$. Hereinafter, $\bm{z}^{\prime}$ specifically indicates that the latent variables originate from the generated images $\hat{\bm{x}}$. Furthermore, the decoder attempts to generate realistic reconstructions $\hat{\bm{x}}$ based on latent variables $\bm{z}$ such that the encoder mistakenly assigns a small ${L}_\mathrm{REG}$ value to the generated images. The regularization term (\ref{eq:regularization_term}) is changed as follows:
\begin{equation}
 \begin{alignedat}{4}
  {L}_\mathrm{REG(encoder)} &= {L}_\mathrm{REG}(E(\bm{x})) + [m - {L}_\mathrm{REG}(E(D(\bm{z})))]^{+} \\
         &= {L}_\mathrm{REG}(E(\bm{x})) + [m - {L}_\mathrm{REG}(E(\hat{\bm{x}}))]^{+}\\
         &= {L}_\mathrm{REG}(\bm{z}) + [m - {L}_\mathrm{REG}(\bm{z}^{\prime})]^{+}\\
         &= {L}_\mathrm{REG}(\bm{z}) + {L}_\mathrm{Margin}(\bm{z}^{\prime}),
 \end{alignedat}
\end{equation}
and
\begin{equation} 
{L}_\mathrm{REG(Decoder)} = {L}_\mathrm{REG}(\bm{z}^{\prime}),
\end{equation} 
where $[\cdot]^{+} = \max(0, \cdot)$, $m$ is a scalar of the positive margin, and ${L}_\mathrm{Margin}(\bm{z}^{\prime}) = [m - {L}_\mathrm{REG}(\bm{z}^{\prime})]^{+}$. When ${L}_\mathrm{Margin}(\bm{z}^{\prime})$ is positive, the equations above prompt a min--max game between the encoder and decoder. Finally, the overall objectives are described as follows:
\begin{equation} 
{L}_\mathrm{Encoder} = {L}_\mathrm{REG}(\bm{z}) + \alpha {L}_\mathrm{Margin}(\bm{z}^{\prime}) + \beta {L}_\mathrm{AE},
\end{equation} 
\begin{equation} 
{L}_\mathrm{Decoder} = \alpha {L}_\mathrm{REG}(\bm{z}^{\prime}) + \beta {L}_\mathrm{AE},
\end{equation} 
where $\alpha$ and $\beta$ are weighting parameters that balance the importance of each loss term. 

\subsubsection{Latent Representation Searching}

We further applied latent representation searching to improve the latent representation for achieving a better reconstruction (Fig. \ref{fig:image_reconstruction_networks}b), as inspired by the method used in AnoGAN \cite{Schlegl2017}. To obtain the optimal latent representation $\bm{z}_{*}$, the method begins with the first latent position $\bm{z}_{1}$, which is initially mapped by the encoder capturing the input images $\bm{x}$. Subsequently, $\bm{z}_{1}$ is input into the decoder to yield the first reconstructed images $\hat{\bm{x}}_{1}$. Using the reconstruction error ${L}_\mathrm{AE}$ for the residual between $\bm{x}$ and $\hat{\bm{x}}_{1}$, the latent representation can be updated based on the gradients for objective minimization, shifting $\bm{z}_{1}$ to a better position $\bm{z}_{2}$ in the latent space. Subsequently, the deviation of the secondary reconstructed images $\hat{\bm{x}}_{2}$ with respect to $\bm{x}$ is evaluated for the next objective. This update rule is described as an iterative process from $\bm{z}_{i}$ to $\bm{z}_{i+1}$ to minimize ${L}_{AE}(\bm{x}, \hat{\bm{x}}_{i})$. After sufficient training steps of this optimization process, we can expect $\bm{z}_{*}$ to yield a better image reconstruction $\bm{\hat{x}}_{*}$.

\subsection{Discriminative Network}

\begin{figure}
  \centering
  \includegraphics[width=0.8\hsize]{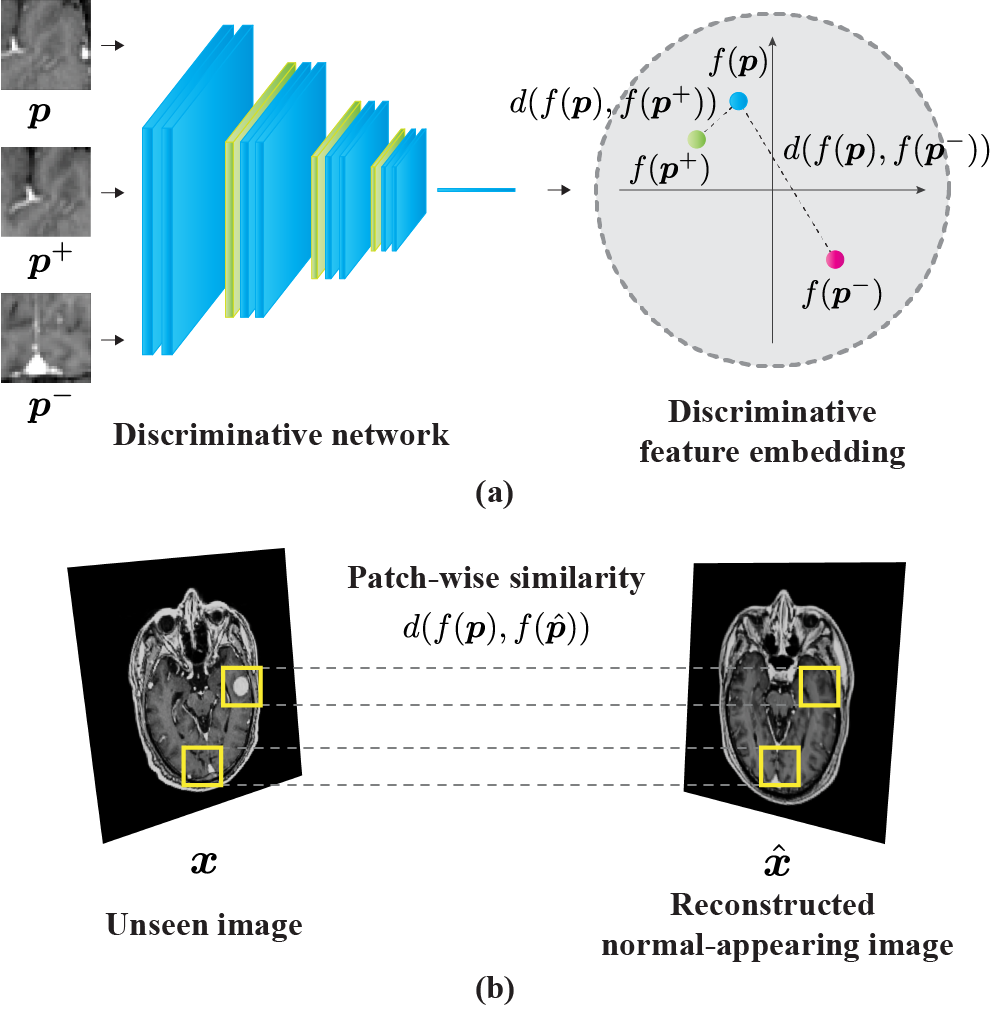}
  \caption{\textbf{Discriminative networks for recognizing local patterns of normal brain anatomy.} (a) Discriminative networks learn patch-wise discriminative embeddings based on metric learning techniques using triplet margin loss. (b) By calculating the patch-wise similarity in discriminative embeddings between unseen images and reconstructed normal-appearing images, the abnormality distribution can be measured as abnormality scores.}
  \label{fig:discriminative_networks}
\end{figure}

Whereas the reconstruction network learns the normal anatomy of the entire image, the discriminative network learns the pattern of the patch-level normal appearance based on metric learning. The discriminative network is essential for detecting abnormalities by comparing query images with reconstructed normal-appearing images. 

At the training stage, for every randomly cropped patch $\bm{p}$ from the normal images, a triplet of patches $(\bm{p}, \bm{p}^+, \bm{p}^-)$ is prepared (Fig. \ref{fig:discriminative_networks}a). Positive patches $\bm{p}^+$ are obtained from small random affine translations around the original patch $\bm{p}$ and small changes in the image intensity and range. Negative patches $\bm{p}^-$ are created by random cropping from the same images. The discriminative network $f$ learns discriminative embeddings using the triplet margin loss \cite{BMVC2016_119} as follows:
\begin{equation} 
\label{eq:triplet_margin_loss}
L_\mathrm{Margin} = \max \{d(f(\bm{p}), f(\bm{p}^+)) - d(f(\bm{p}), f(\bm{p}^-)) + 1, 0\},
\end{equation} 
where $d(\cdot , \cdot)$ is the L2 distance between two embedded features. 

When detecting abnormalities in unseen images, the patch-wise similarity of embedded features between input images $\bm{x}$ and reconstructed images $\hat{\bm{x}}$ is calculated by the reconstruction network using the L2 distance (Fig. \ref{fig:discriminative_networks}b). We considered the similarity measure $d(f(\bm{p}), f(\hat{\bm{p}}))$ as the \emph{abnormality score} of the center pixel in each patch. The distribution of abnormality scores per image was standardized using Z-score normalization. As the reconstruction network reproduces the normal-appearing image more faithfully, the contrast with the abnormal part in the unseen image becomes clearer, and the detection performance using abnormality score can be expected to improve. 


\subsection{Anatomical Fidelity}
\label{sec:anatomical_fidelity}

We propose the concept of \emph{anatomical fidelity} to quantitatively evaluate the reconstruction. The generated images should exhibit both high-resolution details at local points and anatomical consistency between distant portions of the image. To assess the anatomical fidelity, we decompose this index into two measures, i.e., the \emph{quality score} and \emph{overlap score}, by exploiting a multiclass segmentation network for normal anatomical classes trained on the same image domain. 


\subsubsection{Quality Score}
The quality score reflects the high resolution of the generated images. Generally, an imperfect image generation yields blurry results with no sharp details, and such images can be regarded as out-of-distribution samples by a segmentation network trained on a dataset from the same domain \cite{hendrycks2016baseline}. Therefore, we used the entropy of the softmax distribution of the segmentation network as a quality measure of the generated images. The quality score is defined as follows:
\begin{equation} 
S_\mathrm{quality} = \mathrm{Entropy}(\bm{x}) - \mathrm{Entropy}(\hat{\bm{x}}),
\end{equation} 
where $\mathrm{Entropy}(\bm{x}) = - \sum_{y_k} \sum_{i,j} p(y_k|{x}_{i,j}) \log{p(y_k|{x}_{i,j}})$ and $p(y_k|{x}_{i,j})$ is the conditional probability that an $i,j$-th pixel $x_{i,j}$ belongs to class $y_k$. 

\subsubsection{Overlap Score}

If the generated images $\hat{\bm{x}}$ are geometrically well-aligned with the input images $\bm{x}$, the anatomical classes of these images, as predicted by the segmentation network, should overlap broadly. Based on this assumption, we used the Dice score, which is a popular overlap measure for segmentation \cite{Sudre2017GeneralisedDO}, to quantify the similarity with respect to each anatomical class as follows:
\begin{equation} 
S_\mathrm{overlap}(\bm{x}, \hat{\bm{x}}) = \frac{1}{k} \sum_{k \in \mathcal{K}} \frac{2|y(\bm{x})_{k} \cap y(\hat{\bm{x}})_{k}|}{|y(\bm{x})_{k}| + |y(\hat{\bm{x}})_{k}|},
\end{equation} 
where $y$ denotes the segmentation output of the segmentation network, subscript $k$ indicates the segmentation map of each label, and $\mathcal{K}$ denotes a set of class indices that appear in the input image. 

\section{Experiments}
\label{sec:experiments}

\begin{figure}
  \centering
  \includegraphics[width=0.8\hsize]{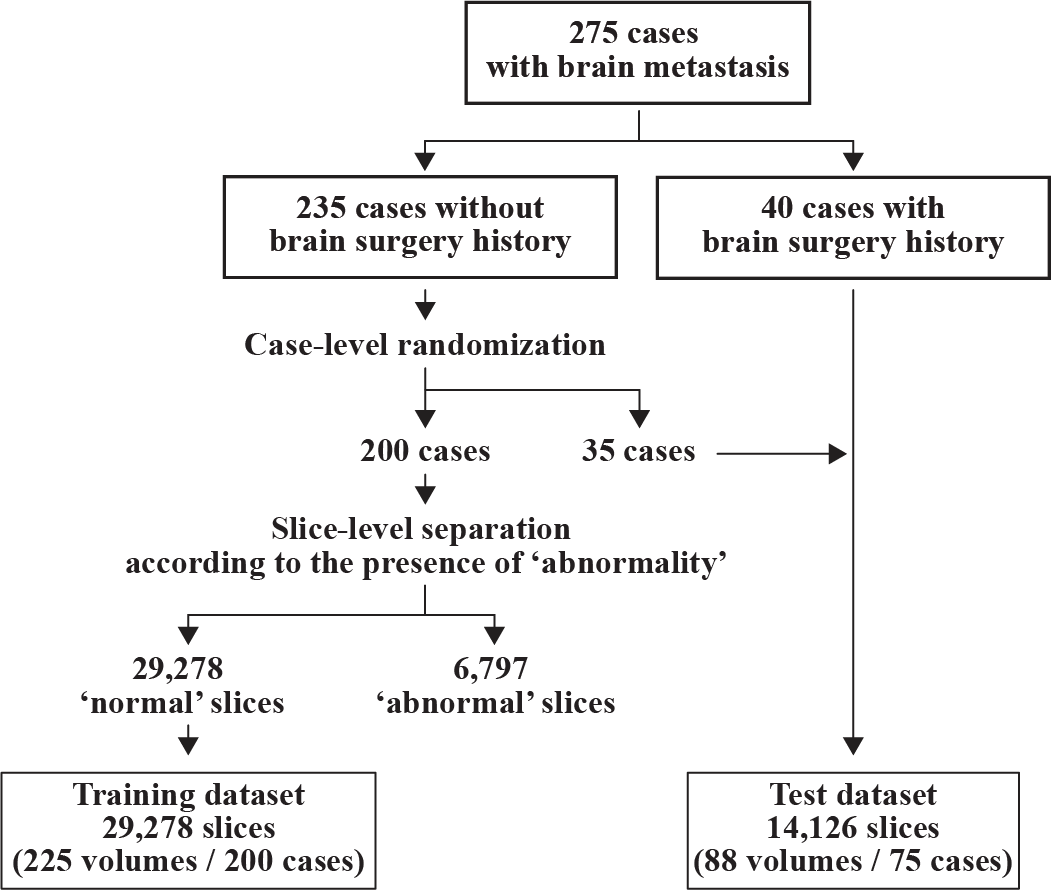}
  \caption{\textbf{Dataset splitting.} In total, 275 cases were included in the present study. From 235 cases with no history of brain surgery, 200 cases were randomly selected, and the data were separated into 36,075 axial slices. Each slice was independently grouped based on the presence of any abnormality from the four classes. Among those slices, 29,278 slices with no annotated abnormalities were assigned to the training dataset. The remaining 35 patients with no history of brain surgery and 40 patients with a history of surgery were integrated into a test dataset.}
  \label{fig:dataset_splitting}
\end{figure}

\subsection{Dataset}
This retrospective, single-center study was approved by our institutional review board. We randomly collected data for 275 patients who underwent MRI analysis for the treatment planning of stereotactic radiotherapy or radiosurgery for brain metastasis using CyberKnife (Accuray Inc., Sunnyvale, CA) during a particular period in our institution. In some cases, several MRI analyses were performed during the study period; therefore, the dataset contained a total of 313 MRI volumes. All the acquired MRI volumes contained at least one metastatic brain lesion. The imaging protocol acquired contrast-enhanced three-dimensional gradient-echo (CE3D-GRE) sequences with spatial resolutions of $1 \times 1 \times 1\ \mathrm{mm}^3$ or higher. 

For all 313 MRI volumes, slice-by-slice ground truth segmentation was performed by an experienced radiation oncologist, who manually delineated the regions of interest for four classes: \emph{metastatic brain tumor}, \emph{extracranial metastatic tumor}, \emph{postoperative cavity}, and \emph{structural change, NOS}. In particular, the structural change, NOS class encompassed any other gross structural changes that occurred in the brain parenchyma; it was included to enhance the comprehensiveness of the dataset. For example, ischemic changes due to a previous stroke were included in this category. It is noteworthy that minor postoperative changes or deformations that occurred in other anatomies outside the brain parenchyma, such as the skull or subcutaneous tissue, were not delineated because a clear boundary definition was difficult to obtain. In addition, edematous changes in the brain parenchyma were recognized in some cases but were not categorized because of their unclear boundaries in the CE3D-GRE sequence.


\subsection{Pre-processing}
\label{sec:pre-processing}

First, all 3D MRI volumes $\bm{X}$ were resampled to a voxel size of $1 \times 1 \times 1\, \mathrm{mm}^{3}$ using cubic interpolation. Subsequently, intensity normalization based on histogram matching was applied to normalize the intensity variations, where the white matter peak was mapped to the middle value of the subdivided points. Next, each 3D MRI volume $\bm{X}$ was decomposed into a collection of two-dimensional (2D) slices $\{\bm{x}_1, \bm{x}_2, \cdots, \bm{x}_k\}$ and randomly shuffled. Every 2D slice was center cropped to a size of $256 \times 256$. During training, data augmentation (including horizontal flipping, random scaling, and rotation) was performed. Finally, every image was renormalized to the range $[-1, 1]$.


\subsection{Implementation}

All the experiments were implemented in Python 3.7 with PyTorch library 1.2.0 \cite{NEURIPS2019_9015}, using an NVIDIA Tesla V100 graphics processing unit and CUDA 10.0. For all the networks, Adam optimization \cite{kingma2014adam} was used for the training. Network initialization was performed using the method described in \cite{he2015delving}.

\subsubsection{Image Reconstruction Network Using VAE}

The encoder comprised residual blocks \cite{He2015}, in which two [{\it convolution} + {\it batch normalization} + {\it LeakyReLU}] sequences were processed with residual connections. From the first to the last residual block, the encoder utilized $32-64-128-256-512-512$ filter kernels. Each residual block was followed by an average pooling function to halve the feature map size. The input was required to be a grayscale 2D image of size $1 \times 256 \times 256$ $(= \mathrm{channel} \times \mathrm{height} \times \mathrm{width})$ with a normalized range of $[-1, 1]$, and the output was designed to be two individual variables, $\bm{\mu}$ and $\bm{\sigma}$, having the same size as $\bm{z}$, i.e., $128 \times 4 \times 4$. 

The decoder architecture was almost symmetrical to that of the encoder. From the first to the last residual block, the decoder utilized $512-512-256-128-64-32-16$ filter kernels. The residual blocks comprised two [{\it convolution} + {\it batch normalization} + {\it LeakyReLU}] sequences, followed by an upsampling layer that utilized the interpolation function coupled with a convolutional function to enlarge the feature map size. The latent variables with a size of $128 \times 4 \times 4$ were passed through the decoder to yield reconstructed 2D images measuring $1 \times 256 \times 256$. {\it Tanh} activation was applied to the output to restrict the generated images to the range $[-1, 1]$. 

Learning rates of $1 \times 10^{-4}$ and $5 \times 10^{-3}$ were used for the encoder and decoder, respectively. The other hyperparameters were determined as follows: batch size = 120, maximum number of epochs = 200.

\subsubsection{Image Reconstruction Network Using IntroVAE}

The network architecture was the same as that of the VAE. The hyperparameters were determined as follows: batch size = 120, maximum number of epochs = 200, $\alpha = 0.5$, $\beta = 0.04$, and $m = 120$. We added structural similarity \cite{Wang2004} as a constraint to the L2 loss for image reconstruction to capture the perceptual similarity and interdependencies between local pixel regions \cite{Bergmann2018}. 


\subsubsection{Latent Representation Searching}

We updated the latent variables with 100 steps using a learning rate of $1 \times 10^{-3}$. During optimization, all the parameters of the encoder and decoder were fixed. 

\begin{figure*}[t]
  \centering
  \includegraphics[width=0.8\hsize]{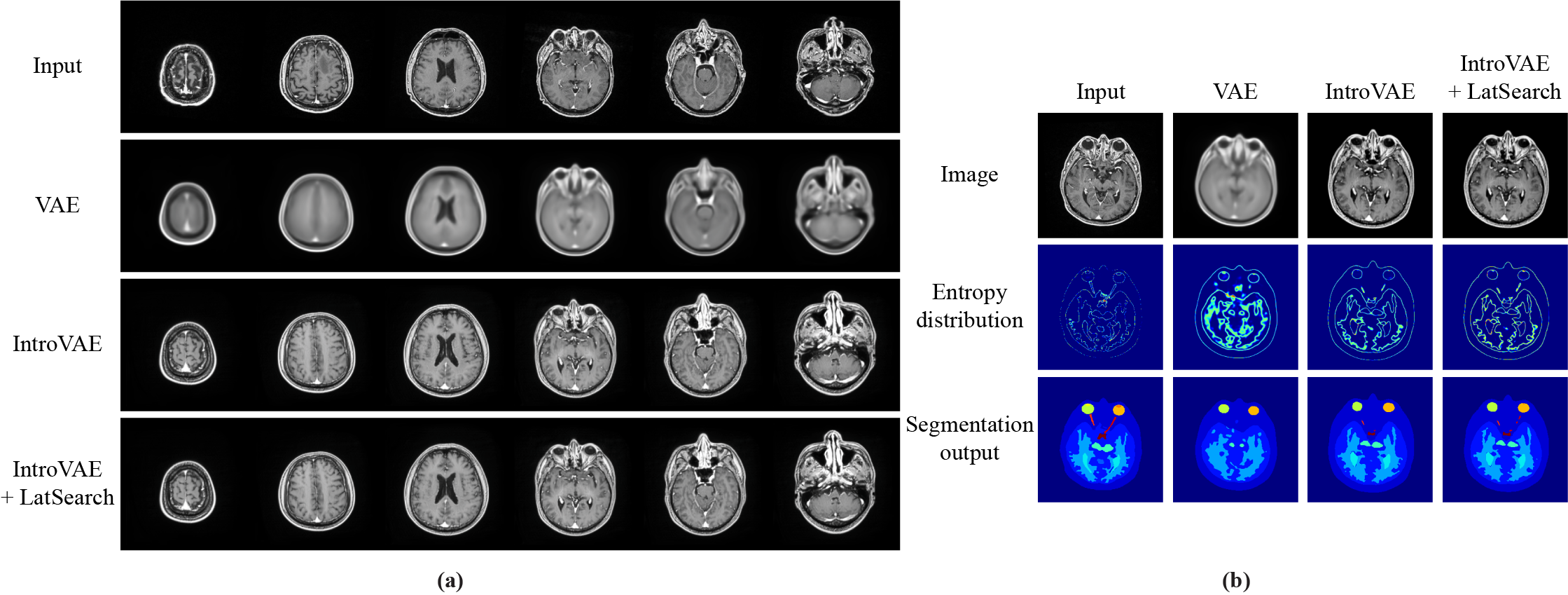}
  \caption{\textbf{Qualitative and quantitative comparison of image reconstruction results of VAE, IntroVAE, and IntroVAE+LatSearch.} (a) Input images, images generated by VAE, IntroVAE, and IntroVAE+LatSearch are shown on the first to the fourth row, respectively. These images confirm that fine details were well reproduced using IntroVAE instead of VAE as the backbone. (b) Pixel-wise softmax entropy provided by the segmentation network is shown in the second row. Higher values appear in obscure regions of the generated images, especially those generated by VAE. As shown in the third row, anatomical consistency can be assessed by calculating the concordance in segmentation labels between the input and reconstructed images.}
  \label{fig:reconstruction_results}
\end{figure*}

\subsubsection{Discriminative Network}

The discriminative network comprised two sequential residual blocks \cite{He2015}, in which two [{\it convolution} + {\it batch normalization} + {\it ReLU}] sequences were processed with residual connections. From the first to the last block, the network utilized $64-128-256-512$ filter kernels. A max pooling function was inserted after each block to halve the size of the feature maps. After the last block, a multilayer perceptron comprising one hidden layer with a size of $1024$ generated a discriminative embedding vector, whose dimensional size was set to $256$. The size of the input patch was set to $1 \times 32 \times 32$ $(= \mathrm{channel} \times \mathrm{height} \times \mathrm{width})$.

\subsubsection{Segmentation Network for Anatomical Fidelity}
\label{sec:segmentation_network_for_anatomical_fidelity}

We prepared a pre-trained segmentation network that can perform pixel-wise classification for 13 anatomical classes: the \emph{whole brain}, \emph{cerebellum}, \emph{white matter}, \emph{cerebrospinal fluid}, \emph{brain stem}, \emph{pituitary gland}, \emph{right eye}, \emph{right lens}, \emph{left eye}, \emph{left lens}, \emph{right optic nerve}, \emph{left optic nerve}, and \emph{optic chiasm}. These labels were semi-automatically generated using the MultiPlan treatment planning system (Accuray Inc., Sunnyvale, CA) and visually reviewed by an expert radiation oncologist. The dataset comprised 100 MRI volumes that were scanned based on the same protocol but collected independently from the dataset for the main experiment. Each MRI volume was subjected to pre-processing---voxel size normalization, intensity normalization, separation into 2D slices, and center cropping (see Section \ref{sec:pre-processing}). The dataset was divided as follows: 70\% for training (70 volumes), 15\% for validation (15 volumes), and 15\% for testing (15 volumes). The configuration that yielded the best performance on the 15 volumes for validation was obtained as follows. The segmentation network utilized the ResUNet architecture \cite{resunet2020}, and the loss function comprised soft Dice \cite{Sudre2017GeneralisedDO} and focal losses \cite{focal2017}. The learning rate, weight decay, and batch size were set to $1 \times 10^{-5}$, $1 \times 10^{-5}$, and 100, respectively. The model was trained for 1,000 epochs with random rotation for data augmentation. Subsequently, the segmentation performance was evaluated using the 15 volumes reserved for testing.

\subsubsection{Detection Performance Evaluation}
\label{sec:detection_performance_evaluation}
We discovered a significant foreground--background imbalance caused by the limited space of the body within the overall MRI volume; this resulted in a tendency to overestimate the detection performance when the ROC curve was used naively based on all pixels in the images. In fact, as the black background surrounding the body did not provide any information for radiologists, it was less emphasized. To evaluate the detection performance fairly, we excluded the background from the calculation. 

Based on these considerations, we evaluated the class-wise detection performance for the four abnormality classes in the test dataset. The standardized abnormality scores were divided into 1,000 operating thresholds. \emph{True} and \emph{false positives} were determined at the voxel level, i.e., we considered a voxel to be truly positive when its abnormality score exceeded the threshold and overlapped the ground truth annotation, and vice versa. The datasets had multiple class labels for a single abnormality score; therefore, a positive voxel was considered valid if it corresponded with the \emph{any} class label. Subsequently, to plot the class-wise ROC curve, the true positive and false positive rates across the area inside the body were calculated at all operating points for each abnormal class. 

\section{Results}
\label{results}

Hereinafter, \emph{VAE} and \emph{IntroVAE} indicate the proposed reconstruction networks developed using the VAE and IntroVAE as the backbone, respectively. \emph{IntroVAE+LatSearch} represents the reconstruction networks based on IntroVAE, followed by latent representation searching.

\subsection{Radiological Characteristics of Abnormality Labels}

The most essential but hidden factor in the comparison of model performance may be the intraclass variability of the target disease \cite{Oakden-hidden}. Therefore, to quantitatively describe the radiological characteristics of the abnormality labels, we calculated the number of lesions and volumes (mean $\pm$ standard deviation). In the test dataset, we discovered $270$ metastatic brain tumors (volume: 2.2 $\pm$ 6.2 mL), $17$ extracranial metastatic tumors (volume range: 4.3 $\pm$ 4.0 mL), $33$ postoperative cavities (volume range: 20.8 $\pm$ 34.1 mL), and $5$ structural changes (NOS) (volume range: 4.1 $\pm$ 7.0 mL).

\subsection{Training Results of Image Reconstruction Networks}

Fig. \ref{fig:reconstruction_results}a shows the sample training results for image reconstruction based on the training dataset. The first row shows the input images, and the second, third, and forth rows show the reconstructed images obtained using VAE, IntroVAE, and IntroVAE+LatSearch, respectively. Even after sufficient training steps, the images generated using VAE tended to be blurry, and the details of the input images were not reproduced. Meanwhile, IntroVAE and IntroVAE+LatSearch generated more realistic-appearing images with detailed structural information. 

To assess the anatomical fidelity of the reconstructed images, we trained a segmentation network for 13 anatomical classes. The segmentation performance of the segmentation network based on the Dice score was as follows: $0.85$ for the entire brain, $0.93$ for the cerebellum, $0.85$ for the white matter, $0.88$ for the cerebrospinal fluid, $0.92$ for the brain stem, $0.70$ for the pituitary gland, $0.85$ for the right eye, $0.71$ for the right lens, $0.89$ for the left eye, $0.59$ for the left lens, $0.60$ for the right optic nerve, $0.60$ for the left optic nerve, and $0.65$ for the optic chiasm.

The pixel-wise softmax entropy was acquired from the segmentation network to evaluate the quality score. As demonstrated in Fig. \ref{fig:reconstruction_results}b, the images generated by the VAE were blurrier than those yielded by IntroVAE, thereby generating disturbed segmentation outputs in accordance with the localization of higher entropy values. The mean $\pm$ standard deviations of the quality score were $-1645.1 \pm 420.0$, $-655.8 \pm 253.9$, and $-617.0 \pm 239.3$ for the VAE, IntroVAE, and IntroVAE+LatSearch, respectively, showing the highest value for IntroVAE+LatSearch.

To calculate the overlap score, the segmentation results for the input and reconstructed images were compared for each anatomical label. The mean $\pm$ standard deviations of the overlap scores were $0.56 \pm 0.09$, $0.558 \pm 0.08$, and $0.59 \pm 0.09$ for the VAE, IntroVAE, and IntroVAE+LatSearch, respectively. The highest value of the overlap score was also observed for IntroVAE+LatSearch. 

\subsection{Distribution of Abnormality Scores}

\begin{figure}
  \centering
  \includegraphics[width=0.8\hsize]{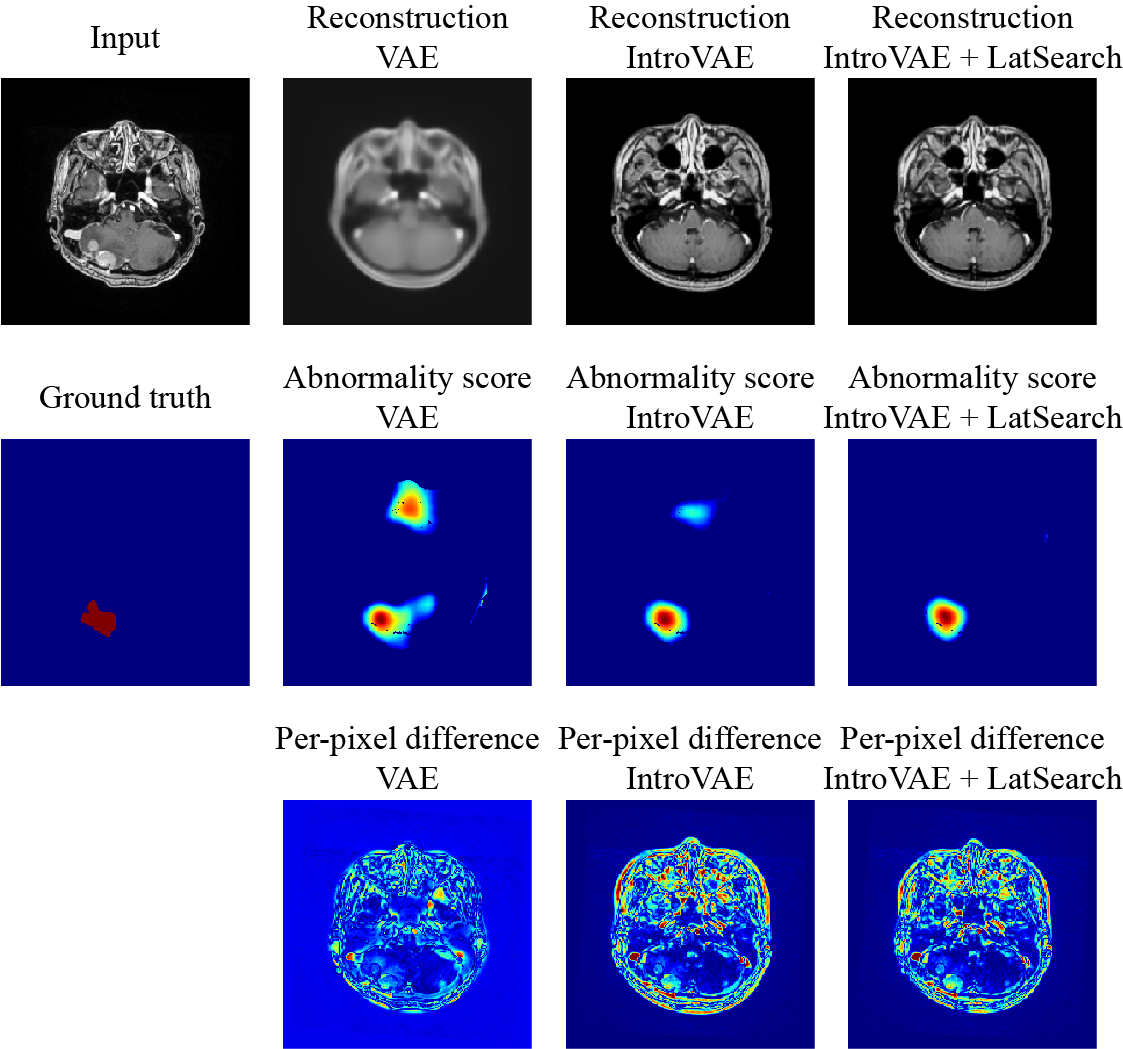}
  \caption{\textbf{Difference between abnormality scores and per-pixel distance of intensities.} Note that the per-pixel L1 distance was extremely sensitive for indistinct image differences, whereas the abnormality score generally reflects the presence of a semantic object (see the region of metastatic brain tumor indicated by the red label). In particular, the accumulation of abnormal scores correlated well with the lesion when using IntroVAE+LatSearch.}
  \label{fig:distance_measurement}
\end{figure}

The previously proposed approaches that utilize image reconstruction networks trained solely on healthy images primarily evaluate pixel-wise residuals by calculating the L1 distances \cite{BAUR2021101952}. Therefore, we compared the per-pixel difference between the input and reconstructed images based on the abnormality score calculated using discriminative embeddings. As shown in Fig. \ref{fig:distance_measurement}, the L1 distance of the pixel intensities yielded a low signal-to-noise ratio. In contrast, a more meaningful distribution of abnormality scores corresponding to the presence of semantic objects (i.e., metastatic brain tumor shown by the red ground-truth label) was observed, where the tendency became the clearest when using IntroVAE+LatSearch. 

\subsection{Detection Performance}

\begin{figure}
  \centering
  \includegraphics[width=0.8\hsize]{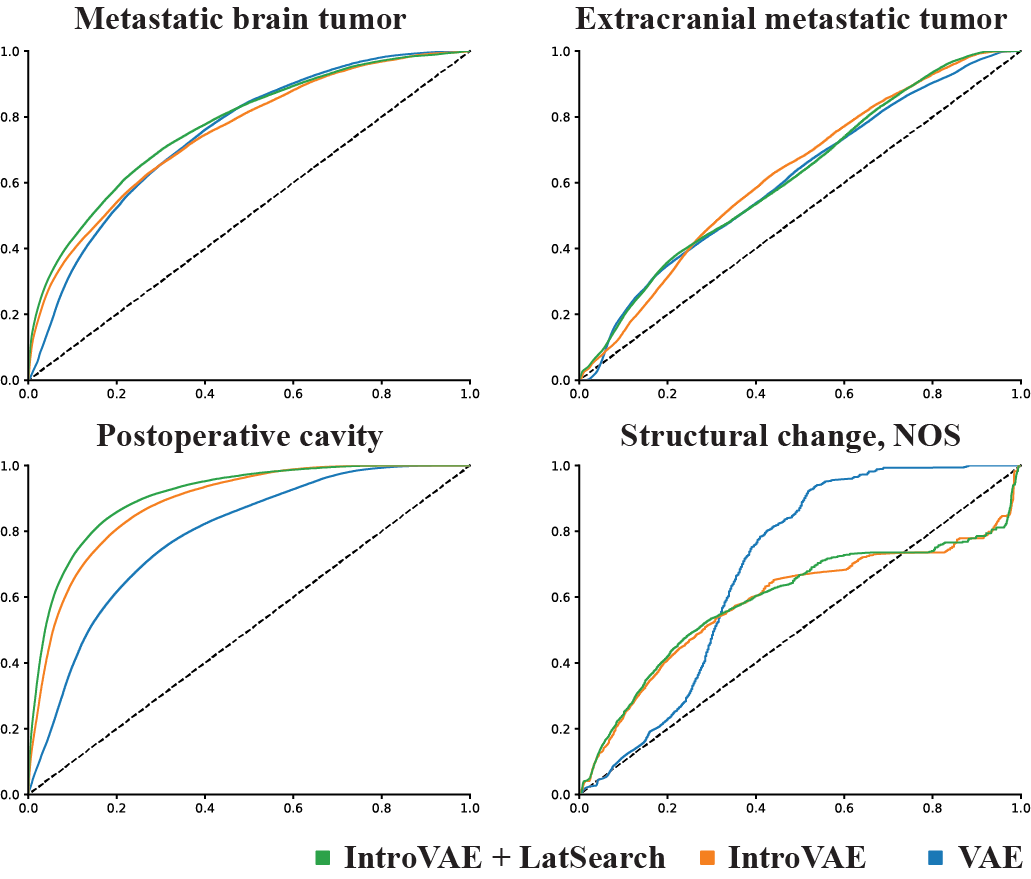}
  \caption{\textbf{Rectified class-wise detection performance for four abnormality classes.} The mean ROC curves were calculated by considering the voxels inside the body to rectify the significant foreground--background imbalance. For each plot, the vertical and horizontal axes indicate true positive and false positive, respectively.}
  \label{fig:roc_curves}
\end{figure}

\begin{table*}[]
\centering
\resizebox{0.75\linewidth}{!}{%
\begin{tabular}{@{}lllllll@{}}
\toprule
\textbf{Abnormal class}                                 & \textbf{Reconstruction networks} & \textbf{ROC--AUC} & \textbf{Sensitivity} & \textbf{Specificity} & \textbf{Precision} & \textbf{F1 score} \\ \midrule
\multirow{3}{*}{\textbf{Metastatic brain tumor}}        & \textbf{VAE}                     & 0.75 $\pm$ 0.16        & 0.88 $\pm$ 0.12            & 0.89 $\pm$ 0.06            & 0.007 $\pm$ 0.012        & 0.014 $\pm$ 0.024       \\
                                                        & \textbf{IntroVAE}                & 0.76 $\pm$ 0.17        & 0.87 $\pm$ 0.11            & 0.89 $\pm$ 0.07            & 0.011 $\pm$ 0.021        & 0.021 $\pm$ 0.038       \\
                                                        & \textbf{IntroVAE+LatSearch}      & 0.78 $\pm$ 0.17        & 0.86 $\pm$ 0.12            & 0.90 $\pm$ 0.07            & 0.014 $\pm$ 0.025        & 0.027 $\pm$ 0.047       \\ \midrule
\multirow{3}{*}{\textbf{Extracranial metastatic tumor}} & \textbf{VAE}                     & 0.63 $\pm$ 0.21        & 0.93 $\pm$ 0.08            & 0.82 $\pm$ 0.09            & 0.003 $\pm$ 0.003        & 0.007 $\pm$ 0.006       \\
                                                        & \textbf{IntroVAE}                & 0.61 $\pm$ 0.21        & 0.92 $\pm$ 0.06            & 0.81 $\pm$ 0.09            & 0.005 $\pm$ 0.006        & 0.009 $\pm$ 0.012       \\
                                                        & \textbf{IntroVAE+LatSearch}      & 0.61 $\pm$ 0.24        & 0.92 $\pm$ 0.06            & 0.81 $\pm$ 0.10            & 0.006 $\pm$ 0.009        & 0.011 $\pm$ 0.017       \\ \midrule
\multirow{3}{*}{\textbf{Postoperative cavity}}          & \textbf{VAE}                     & 0.79 $\pm$ 0.14        & 0.86 $\pm$ 0.18            & 0.89 $\pm$ 0.07            & 0.021 $\pm$ 0.029        & 0.040 $\pm$ 0.051       \\
                                                        & \textbf{IntroVAE}                & 0.89 $\pm$ 0.09        & 0.87 $\pm$ 0.17            & 0.93 $\pm$ 0.04            & 0.039 $\pm$ 0.061        & 0.068 $\pm$ 0.098       \\
                                                        & \textbf{IntroVAE+LatSearch}      & 0.91 $\pm$ 0.08        & 0.88 $\pm$ 0.17            & 0.94 $\pm$ 0.04            & 0.048 $\pm$ 0.073        & 0.083 $\pm$ 0.112       \\ \midrule
\multirow{3}{*}{\textbf{Structural change, NOS}}        & \textbf{VAE}                     & 0.68 $\pm$ 0.11        & 0.92 $\pm$ 0.06            & 0.85 $\pm$ 0.02            & 0.004 $\pm$ 0.007        & 0.008 $\pm$ 0.013       \\
                                                        & \textbf{IntroVAE}                & 0.59 $\pm$ 0.35        & 0.93 $\pm$ 0.07            & 0.82 $\pm$ 0.09            & 0.005 $\pm$ 0.007        & 0.009 $\pm$ 0.015       \\
                                                        & \textbf{IntroVAE+LatSearch}      & 0.60 $\pm$ 0.36        & 0.86 $\pm$ 0.12            & 0.84 $\pm$ 0.11            & 0.006 $\pm$ 0.010        & 0.011 $\pm$ 0.019       \\ \bottomrule
\end{tabular}}
\caption{Comparison of detection performance based on image reconstruction networks.} 
\label{tab:detection_performance}
\end{table*}

Fig. \ref{fig:roc_curves} shows the rectified ROC curves, which were calculated by considering the voxels inside the body only. The detection performance (mean $\pm$ standard deviation) based on the rectified ROC--AUCs is summarized in Table \ref{tab:detection_performance}. It can be seen that metastatic brain tumor and postoperative cavity were detected with a sufficient ROC--AUC exceeding 0.7. Between these two classes, IntroVAE+LatSearch outperformed the other two image reconstruction networks, namely VAE and IntroVAE, where the order of the higher detectability was compatible with the higher anatomical fidelity. Meanwhile, we discovered a low precision due to the much higher prevalence of ``normal'' voxels over ``abnormal'' voxels; this can yield a larger number of false-positive voxels. 

\begin{figure}[t]
  \centering
  \includegraphics[width=0.8\hsize]{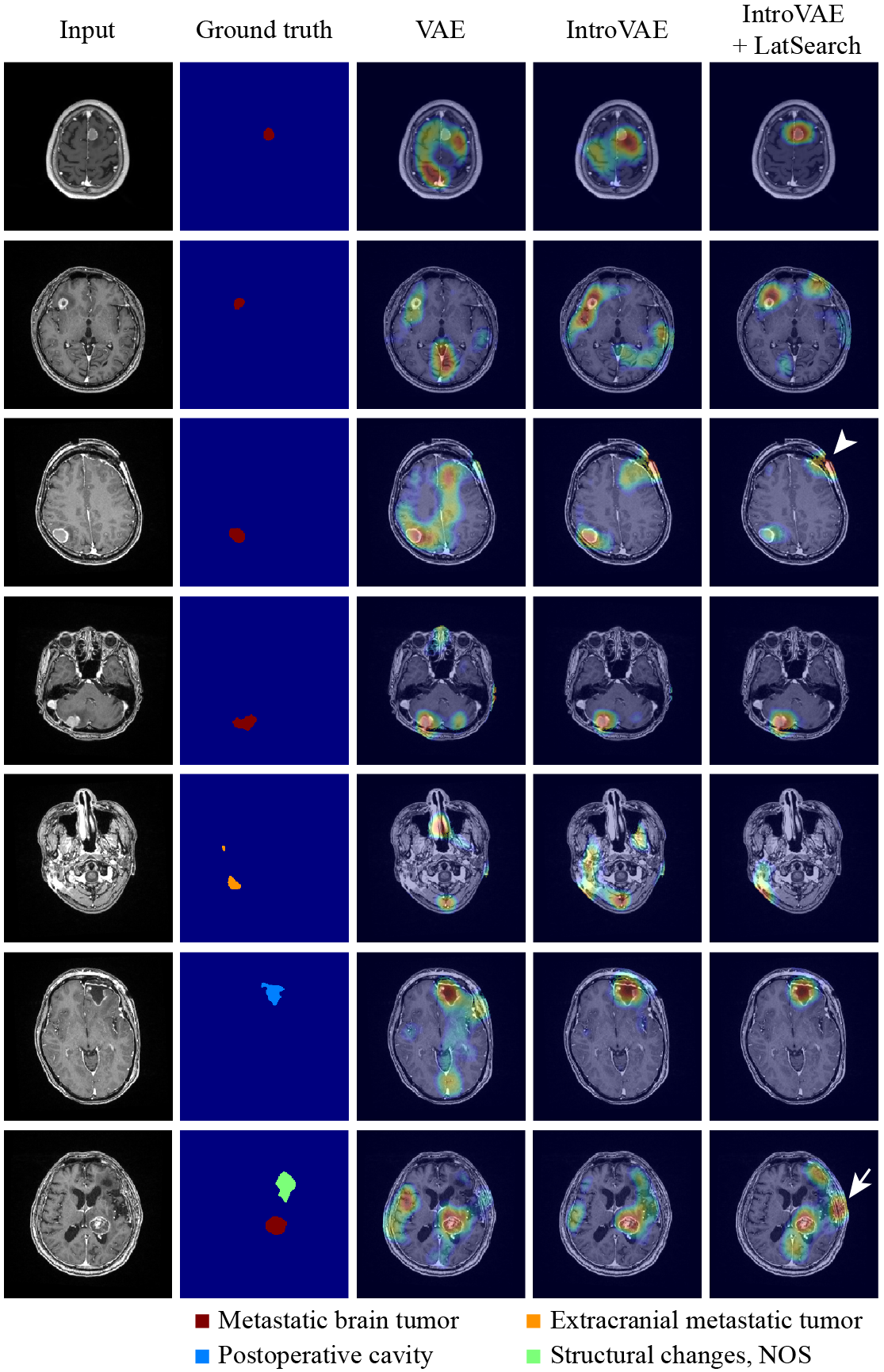}
  \caption{\textbf{Visual examples of reconstructed images, labels, and abnormality scores.} Metastatic brain tumors, extracranial metastatic tumors, and postoperative cavities corresponded to significant accumulation of abnormality scores. Other undefined structural deformations outside the brain parenchyma were detected based on the intensity of the score (arrow and arrowhead).}
  \label{fig:discrimination_results}
\end{figure}

Finally, the visual examples are shown in Fig. \ref{fig:discrimination_results}. When using IntroVAE+LatSearch, metastatic brain tumors, extracranial metastatic tumors, and postoperative cavities were associated with distinct abnormality scores co-localized in the labeled regions. In contrast, other structural changes yielded accumulations of abnormality scores in the labeled region; however, the distribution was not consistent. It is noteworthy that some undefinable structural deformations outside the brain parenchyma were detected based on the intensity of the score, as indicated by the arrow and arrowhead. 

\section{Discussion and Conclusion}
\label{sec:discussion_and_conclusion}

We introduced a novel unsupervised abnormality detection algorithm in brain MRI and evaluated its performance using a comprehensively annotated dataset for structural abnormalities in brain parenchyma. The concept was as follows: If an image reconstruction network can be trained to learn the global ``normal'' brain anatomy, then the ``abnormal'' lesions in unseen images can be identified based on the local differences from those generated as ``normal'' by a discriminative network. To the best of our knowledge, this is the first study pertaining to the unsupervised abnormality detection of medical imaging using metric learning with a synthetic normal image as a reference. As shown in Fig. \ref{fig:distance_measurement}, the per-pixel difference using the L1 distance yielded a low signal-to-noise ratio owing to the small intensity differences in insignificant details. In contrast, the distance measurement based on the discriminative embeddings yielded abnormality scores that corresponded semantically with the presence of abnormal classes. Thus, by combining reconstruction-based approaches and discriminative-boundary-based approaches from two perspectives (i.e., global and local features of normal anatomy of medical images), we demonstrated a notable performance of the abnormality detection framework trained without labels. 


Furthermore, we devised a concept of anatomical fidelity to quantitatively evaluate the generated medical images. To assess the visual quality and validity of the generated images, an intuitive method using human annotators for the evaluation was utilized; however, it is not a feasible method because of the associated costs. Hence, the inception score \cite{salimans2016improved} and Fr\'echet inception distance \cite{heusel2017gans} were introduced. However, these indices were derived from a model trained based on ImageNet containing a large set of natural images. Therefore, they cannot be applied straightforwardly to generative models based on datasets of other domains \cite{barratt2018note}. We confirmed that the best anatomical fidelity can be achieved using IntroVAE+LatSearch. This trend was generally consistent with the overall detection performance, especially when focusing on the two abnormality classes (i.e., metastatic brain tumor and postoperative cavity).

In contrast to previous studies \cite{BAUR2021101952}, we did not use any regional extraction techniques such as skull-stripping as a pre-process in our experiment. Moreover, the abnormality classes such as metastatic brain tumors were small in volume, rendering them difficult targets to detect. Considering these difficulties, the detection performance of ROC--AUC exceeding 0.7 in metastatic brain tumor and postoperative cavity can be promising. In addition, even though the performance was moderate in comparison with those of supervised frameworks for brain metastasis \cite{Cho2020}, information regarding whether the reported values were rectified to exclude the region outside the body, which can cause overestimation, was not available. For extracranial metastatic tumors, we assumed that the low ROC--AUC may be caused by the much lower frequency of the extracranial region compared with that of brain parenchyma, rendering it difficult for the discriminative network to learn discriminative embeddings. In addition, the number of regions labeled as structural change, NOS was extremely small for interpreting their detectability.


The proposed method is generalizable and can be easily extended to other detection applications. We adopted VAE-based reconstruction networks because the network architecture can be simple, featuring an encoder--decoder network pair; however, it is noteworthy that our framework, which utilizes a discriminative network, can be used with other generative models such as GANs \cite{BAUR2021101952}. In the future, we will develop unsupervised abnormality detection models that will benefit real clinical scenarios.

\bibliographystyle{IEEEtran}
\bibliography{reference}

\begin{thebibliography}{10}
\providecommand{\url}[1]{#1}
\csname url@samestyle\endcsname
\providecommand{\newblock}{\relax}
\providecommand{\bibinfo}[2]{#2}
\providecommand{\BIBentrySTDinterwordspacing}{\spaceskip=0pt\relax}
\providecommand{\BIBentryALTinterwordstretchfactor}{4}
\providecommand{\BIBentryALTinterwordspacing}{\spaceskip=\fontdimen2\font plus
\BIBentryALTinterwordstretchfactor\fontdimen3\font minus
  \fontdimen4\font\relax}
\providecommand{\BIBforeignlanguage}[2]{{%
\expandafter\ifx\csname l@#1\endcsname\relax
\typeout{** WARNING: IEEEtran.bst: No hyphenation pattern has been}%
\typeout{** loaded for the language `#1'. Using the pattern for}%
\typeout{** the default language instead.}%
\else
\language=\csname l@#1\endcsname
\fi
#2}}
\providecommand{\BIBdecl}{\relax}
\BIBdecl

\bibitem{Alexnet_NIPS2012_4824}
A.~Krizhevsky, I.~Sutskever, and G.~E. Hinton, ``Imagenet classification with
  deep convolutional neural networks,'' in \emph{Advances in Neural Information
  Processing Systems 25 (NeurIPS)}, 2012, pp. 1097--1105.

\bibitem{Chalapathy2019}
R.~Chalapathy and S.~Chawla, ``Deep learning for anomaly detection: A survey,''
  \emph{arXiv preprint arXiv:1901.03407}, 2019.

\bibitem{Kingma2013}
D.~P. Kingma and M.~Welling, ``Auto-encoding variational bayes,'' in \emph{The
  2nd International Conference on Learning Representations (ICLR)}, 2014.

\bibitem{Rezende2014}
D.~J. Rezende, S.~Mohamed, and D.~Wierstra, ``Stochastic backpropagation and
  approximate inference in deep generative models,'' in \emph{The 31st
  International Conference on Machine Learning (ICML)}, vol.~32, 2014, pp.
  1278--1286.

\bibitem{Huang2018}
H.~Huang, Z.~Li, R.~He, Z.~Sun, and T.~Tan, in \emph{Advances in Neural
  Information Processing Systems 31 (NeurIPS)}, 2018, pp. 52--–63.

\bibitem{BAUR2021101952}
C.~Baur, S.~Denner, B.~Wiestler, S.~Albarqouni, and N.~Navab, ``{Autoencoders
  for unsupervised anomaly segmentation in brain MR images: A comparative
  study},'' \emph{Med Image Anal.}, vol.~69, p. 101952, 2021.

\bibitem{VanLeemput2001}
K.~{Van Leemput}, F.~Maes, D.~Vandermeulen, A.~Colchester, and P.~Suetens,
  ``{Automated segmentation of multiple sclerosis lesions by model outlier
  detection},'' \emph{IEEE Trans Med Imaging.}, vol.~20, no.~8, pp. 677--688,
  2001.

\bibitem{Prastawa2004}
M.~Prastawa, E.~Bullitt, S.~Ho, and G.~Gerig, ``{A brain tumor segmentation
  framework based on outlier detection},'' \emph{Med Image Anal.}, vol.~8,
  no.~3, pp. 275--283, 2004.

\bibitem{Shiee2010}
N.~Shiee, P.~L. Bazin, A.~Ozturk, D.~S. Reich, P.~A. Calabresi, and D.~L. Pham,
  ``{A topology-preserving approach to the segmentation of brain images with
  multiple sclerosis lesions},'' \emph{Neuroimage.}, vol.~49, no.~2, pp.
  1524--1535, 2010.

\bibitem{Weiss2013}
N.~Weiss, D.~Rueckert, and A.~Rao, ``{Multiple sclerosis lesion segmentation
  using dictionary learning and sparse coding},'' in \emph{Medical Image
  Computing and Computer-Assisted Intervention -- MICCAI 2013}, vol. 8149,
  2013, pp. 735--742.

\bibitem{Tomas-Fernandez2015}
X.~Tomas-Fernandez and S.~K. Warfield, ``{A model of population and subject
  (MOPS) intensities with application to multiple sclerosis lesion
  segmentation},'' \emph{IEEE Trans Med Imaging.}, vol.~34, no.~6, pp.
  1349--1361, 2015.

\bibitem{Zeng2016}
K.~Zeng, G.~Erus, A.~Sotiras, R.~T. Shinohara, and C.~Davatzikos,
  ``{Abnormality Detection via Iterative Deformable Registration and
  Basis-Pursuit Decomposition},'' \emph{IEEE Trans Med Imaging.}, vol.~35,
  no.~8, pp. 1937--1951, 2016.

\bibitem{An2015VariationalAB}
J.~An and S.~Cho, ``Variational autoencoder based anomaly detection using
  reconstruction probability,'' in \emph{Seoul National University Data Mining
  Center 2015-2 Special Lecture on IE}, 2015.

\bibitem{Chalapathy2017}
R.~Chalapathy, A.~K. Menon, and S.~Chawla, ``Robust, deep and inductive anomaly
  detection,'' in \emph{Machine Learning and Knowledge Discovery in Databases},
  vol. 10534, 2017, pp. 36--51.

\bibitem{Goodfellow2014}
I.~J. Goodfellow, J.~Pouget-Abadie, M.~Mirza, B.~Xu, D.~Warde-Farley, S.~Ozair,
  A.~Courville, and Y.~Bengio, ``Generative adversarial nets,'' in
  \emph{Advances in Neural Information Processing Systems 27 (NeurIPS)}, 2014,
  pp. 2672--2680.

\bibitem{Schlegl2017}
T.~Schlegl, P.~Seeb{\"{o}}ck, S.~M. Waldstein, U.~Schmidt-Erfurth, and
  G.~Langs, ``Unsupervised anomaly detection with generative adversarial
  networks to guide marker discovery,'' in \emph{The 25th International
  Conference on Information Processing in Medical Imaging (IPMI)}, 2017, pp.
  146--157.

\bibitem{radford2015unsupervised}
A.~Radford, L.~Metz, and S.~Chintala, ``Unsupervised representation learning
  with deep convolutional generative adversarial networks,'' in \emph{The 4th
  International Conference on Learning Representations (ICLR)}, 2016.

\bibitem{Schlegl2019}
T.~Schlegl, P.~Seeb{\"{o}}ck, S.~M. Waldstein, G.~Langs, and
  U.~Schmidt-Erfurth, ``{f-AnoGAN: Fast unsupervised anomaly detection with
  generative adversarial networks},'' \emph{Med Image Anal.}, vol.~54, pp.
  30--44, 2019.

\bibitem{Pawlowski2018MIDL}
N.~Pawlowski, M.~C.~H. Lee, M.~Rajchl, S.~Mcdonagh, E.~Ferrante, K.~Kamnitsas,
  S.~Cooke, S.~Stevenson, A.~Khetani, T.~Newman, F.~Zeiler, R.~Digby, J.~P.
  Coles, D.~Rueckert, D.~K. Menon, V.~F.~J. Newcombe, and B.~Glocker,
  ``{Unsupervised Lesion Detection in Brain CT using Bayesian Convolutional
  Autoencoders},'' in \emph{MIDL Conference book}, 2018.

\bibitem{Chen2018MIDL}
X.~Chen and E.~Konukoglu, ``Unsupervised detection of lesions in brain mri
  using constrained adversarial auto-encoders,'' in \emph{MIDL Conference
  book}, 2018.

\bibitem{Chen2020}
X.~Chen, S.~You, K.~C. Tezcan, and E.~Konukoglu, ``{Unsupervised lesion
  detection via image restoration with a normative prior},'' \emph{Med Image
  Anal.}, vol.~64, p. 101713, 2020.

\bibitem{Chandola2009}
V.~Chandola, A.~Banerjee, and V.~Kumar, ``Anomaly detection: A survey,''
  \emph{ACM Comput. Surv.}, vol.~41, no.~3, pp. 1--58, 2009.

\bibitem{Seebock2019}
P.~Seebock, S.~M. Waldstein, S.~Klimscha, H.~Bogunovic, T.~Schlegl, B.~S.
  Gerendas, R.~Donner, U.~Schmidt-Erfurth, and G.~Langs, ``{Unsupervised
  Identification of Disease Marker Candidates in Retinal OCT Imaging Data},''
  \emph{IEEE Trans Med Imaging.}, vol.~38, no.~4, pp. 1037--1047, 2019.

\bibitem{Alaverdyan2020}
Z.~Alaverdyan, J.~Jung, R.~Bouet, and C.~Lartizien, ``{Regularized siamese
  neural network for unsupervised outlier detection on brain multiparametric
  magnetic resonance imaging: Application to epilepsy lesion screening},''
  \emph{Med Image Anal.}, vol.~60, p. 101618, 2020.

\bibitem{doersch2016tutorial}
C.~Doersch, ``Tutorial on variational autoencoders,'' \emph{arXiv preprint
  arXiv:1606.05908}, 2016.

\bibitem{Larsen2015}
A.~B.~L. Larsen, S.~K. S\o{}nderby, H.~Larochelle, and O.~Winther,
  ``Autoencoding beyond pixels using a learned similarity metric,'' in
  \emph{The 33rd International Conference on Machine Learning (ICML)}, 2016,
  pp. 1558–--1566.

\bibitem{zhao2017infovae}
S.~Zhao, J.~Song, and S.~Ermon, ``Infovae: Information maximizing variational
  autoencoders,'' in \emph{The 33rd AAAI Conference on Artificial Intelligence
  (AAAI)}, 2019, pp. 5885--5892.

\bibitem{BMVC2016_119}
D.~P. Vassileios~Balntas, Edgar~Riba and K.~Mikolajczyk, ``Learning local
  feature descriptors with triplets and shallow convolutional neural
  networks,'' in \emph{Proceedings of the British Machine Vision Conference
  (BMVC)}, 2016, pp. 119.1--119.11.

\bibitem{hendrycks2016baseline}
D.~Hendrycks and K.~Gimpel, ``A baseline for detecting misclassified and
  out-of-distribution examples in neural networks,'' in \emph{The 5th
  International Conference on Learning Representations (ICLR)}, 2017.

\bibitem{Sudre2017GeneralisedDO}
C.~Sudre, W.~Li, T.~Vercauteren, S.~Ourselin, and M.~J. Cardoso, ``Generalised
  dice overlap as a deep learning loss function for highly unbalanced
  segmentations,'' in \emph{International Workshop on Deep Learning in Medical
  Image Analysis International Workshop on Multimodal Learning for Clinical
  Decision Support}, 2017, pp. 240--248.

\bibitem{NEURIPS2019_9015}
A.~Paszke, S.~Gross, F.~Massa, A.~Lerer, J.~Bradbury, G.~Chanan, T.~Killeen,
  Z.~Lin, N.~Gimelshein, L.~Antiga, A.~Desmaison, A.~Kopf, E.~Yang, Z.~DeVito,
  M.~Raison, A.~Tejani, S.~Chilamkurthy, B.~Steiner, L.~Fang, J.~Bai, and
  S.~Chintala, ``Pytorch: An imperative style, high-performance deep learning
  library,'' in \emph{Advances in Neural Information Processing Systems 32
  (NeurIPS)}, 2019, pp. 8024--8035.

\bibitem{kingma2014adam}
D.~P. Kingma and J.~Ba, ``Adam: {A} method for stochastic optimization,'' in
  \emph{The 3rd International Conference on Learning Representations (ICLR)},
  2015.

\bibitem{he2015delving}
K.~He, X.~Zhang, S.~Ren, and J.~Sun, ``Delving deep into rectifiers: Surpassing
  human-level performance on imagenet classification,'' in \emph{2015 IEEE
  International Conference on Computer Vision (ICCV)}, 2015, pp. 1026--–1034.

\bibitem{He2015}
K.~{He}, X.~{Zhang}, S.~{Ren}, and J.~{Sun}, ``Deep residual learning for image
  recognition,'' in \emph{2016 IEEE Conference on Computer Vision and Pattern
  Recognition (CVPR)}, 2016, pp. 770--778.

\bibitem{Wang2004}
{Zhou Wang}, A.~C. {Bovik}, H.~R. {Sheikh}, and E.~P. {Simoncelli}, ``Image
  quality assessment: from error visibility to structural similarity,''
  \emph{IEEE Trans Image Process.}, vol.~13, no.~4, pp. 600--612, 2004.

\bibitem{Bergmann2018}
P.~Bergmann, S.~L{\"o}we, M.~Fauser, D.~Sattlegger, and C.~Steger, ``Improving
  unsupervised defect segmentation by applying structural similarity to
  autoencoders,'' in \emph{The 14th International Joint Conference on Computer
  Vision, Imaging and Computer Graphics Theory and Applications (VISIGRAPP)},
  2019.

\bibitem{resunet2020}
F.~Diakogiannis, F.~Waldner, P.~Caccetta, and C.~Wu, ``Resunet-a: A deep
  learning framework for semantic segmentation of remotely sensed data,''
  \emph{ISPRS J Photogramm and Remote Sens.}, vol.~16, pp. 94--114, 2020.

\bibitem{focal2017}
T.~{Lin}, P.~{Goyal}, R.~{Girshick}, K.~{He}, and P.~{Dollár}, ``Focal loss
  for dense object detection,'' in \emph{2017 IEEE International Conference on
  Computer Vision (ICCV)}, 2017, pp. 2999--3007.

\bibitem{Oakden-hidden}
L.~Oakden-Rayner, J.~Dunnmon, G.~Carneiro, and C.~Re, ``Hidden stratification
  causes clinically meaningful failures in machine learning for medical
  imaging,'' in \emph{Proceedings of the ACM Conference on Health, Inference,
  and Learning (CHIL)}, 2020, pp. 151–--159.

\bibitem{salimans2016improved}
T.~Salimans, I.~Goodfellow, W.~Zaremba, V.~Cheung, A.~Radford, and X.~Chen,
  ``Improved techniques for training gans,'' in \emph{Advances in Neural
  Information Processing Systems 29 (NeurIPS)}, 2016, pp. 2234–--2242.

\bibitem{heusel2017gans}
M.~Heusel, H.~Ramsauer, T.~Unterthiner, B.~Nessler, and S.~Hochreiter, ``Gans
  trained by a two time-scale update rule converge to a local nash
  equilibrium,'' in \emph{Advances in Neural Information Processing Systems 30
  (NeurIPS)}, 2017, pp. 6629–--6640.

\bibitem{barratt2018note}
S.~T. Barratt and R.~Sharma, ``A note on the inception score,'' in \emph{The
  35th International Conference on Machine Learning (ICML) Workshop on
  Theoretical Foundations and Applications of Deep Generative Models}, 2018.

\bibitem{Cho2020}
S.~J. Cho, L.~Sunwoo, S.~H. Baik, Y.~J. Bae, B.~S. Choi, and J.~H. Kim,
  ``{Brain metastasis detection using machine learning: a systematic review and
  meta-analysis},'' \emph{Neuro-Oncol.}, 2020.

\end{thebibliography}

\end{document}